\documentclass[sigconf, nonacm]{acmart}
\usepackage[utf8]{inputenc}
\usepackage{algorithm}
\usepackage{algpseudocode}
\begin{document}
% --- Title ---
\title{DynaQuery: A Self-Adapting Framework for Querying Structured and Multimodal Data}

% --- Author Information ---
% Note: For a single-author paper, this is the correct format.
\author{Aymane Hassini}
\affiliation{%
  \institution{Al Akhawayn University}
  \city{Ifrane}
  \country{Morocco}
}
\email{A.hassini@aui.ma}

% --- Abstract ---
\begin{abstract}
The rise of Large Language Models (LLMs) has accelerated the long-standing goal of enabling natural language querying over complex, hybrid databases. Yet, this ambition exposes a dual challenge: reasoning jointly over structured, multi-relational schemas and the semantic content of linked unstructured assets. To overcome this, we present \textbf{DynaQuery}—a unified, self-adapting framework that serves as a practical blueprint for next-generation ``Unbound Databases.'' At the heart of DynaQuery lies the \textbf{Schema Introspection and Linking Engine (SILE)}, a novel systems primitive that elevates schema linking to a first-class query planning phase. We conduct a rigorous, multi-benchmark empirical evaluation of this structure-aware architecture against the prevalent unstructured Retrieval-Augmented Generation (RAG) paradigm. Our results demonstrate that the unstructured retrieval paradigm is architecturally susceptible to catastrophic contextual failures, such as  \texttt{SCHEMA\_\allowbreak HALLUCINATION}, leading to unreliable query generation. In contrast, our SILE-based design establishes a substantially more robust foundation, nearly eliminating this failure mode. Moreover, end-to-end validation on a complex, newly curated benchmark uncovers a key generalization principle: the transition from pure \textit{schema-awareness} to holistic \textit{semantics-awareness}. Taken together, our findings provide a validated architectural basis for developing natural language database interfaces that are robust, adaptable, and predictably consistent.

\end{abstract}

\maketitle

% --- Artifact Availability ---
\vspace{0.3cm}
\begingroup\small\noindent\raggedright
\textbf{Artifact Availability:}\\
The source code, data, and/or other artifacts for this paper have been made available at: \url{https://github.com/aymanehassini/DynaQuery}.
\endgroup
% --- End Artifact Availability ---

\section{Introduction}
The task of translating natural language questions into executable SQL queries, known as Text-to-SQL, is a cornerstone of modern data accessibility~\cite{katsogiannis23survey}. The advent of Large Language Models (LLMs) has significantly advanced the state-of-the-art~\cite{li23bird}, promising to democratize data access for non-technical users—a long-standing goal of the field~\cite{androutsopoulos95nlidb}. However, this progress has also illuminated a dual challenge inherent in modern data ecosystems. First, users need to query complex, multi-table relational schemas with high fidelity (the structured challenge). Second, with databases increasingly storing pointers to unstructured assets~\cite{Armbrust21Lakehouse}, users need to reason over the semantic content of linked images and documents (the multimodal challenge), a task explored by recent multimodal query systems~\cite{chen23symphony, patel24lotus}.

While this dual challenge is clear, the benchmarks for measuring progress have evolved at different paces. The structured challenge has seen a significant evolution. The landscape was historically driven by benchmarks like Spider~\cite{yu18spider}, which established the difficulty of generating syntactically complex, cross-domain SQL. Recognizing that even this fell short of real-world scenarios, the community has developed next-generation paradigms~\cite{li23bird,lei24spider2}. BIRD~\cite{li23bird} introduces data-level complexities—such as noisy values, external knowledge, and the need for query efficiency—while Spider 2.0~\cite{lei24spider2} introduces workflow-level complexities by evaluating agentic systems. In parallel, the multimodal challenge remains a more nascent but equally critical frontier. A truly universal data interface must bridge this divide between the structured and the unstructured world.

In response to these challenges, leaders in the data systems community have called for "Unbound Databases"~\cite{madden24unbound}---a new generation of systems designed from the ground up to query the full spectrum of the world's data. While this vision provides a powerful architectural roadmap, the concrete engineering challenges of building such a system remain a critical open problem. Foundational work like the 'Generating Impossible Queries' (GIQ) system~\cite{nejjar25giq} demonstrated a novel method for querying linked multimodal data, but its implementation was a proof-of-concept, leaving the challenges of generalization and system-level adaptation unsolved.

To address this gap, we introduce \textbf{DynaQuery}, a unified, self-adapting framework that provides a practical blueprint for the Unbound Database. Our work is motivated by a core principle: before the advanced complexities of modern benchmarks can be consistently met, a system must first master the foundational task of robust schema linking. DynaQuery is designed around this principle, with a novel query planning engine at its core. This work makes the following scientific contributions:

\begin{itemize}
    \item \textbf{A Rigorous Empirical Analysis of Linking Architectures:} We provide a definitive, multi-benchmark comparison of our structure-aware linking engine (SILE) versus unstructured RAG~\cite{lewis20rag}. Our programmatic failure analysis systematically demonstrates that a primary source of RAG's architectural brittleness is \texttt{SCHEMA\_HALLUCINATION}, a catastrophic failure mode that our approach nearly eliminates.

    \item \textbf{An Empirical Analysis of Architectural Trade-offs in Pluggable Decision Modules:} We provide a rigorous analysis of the trade-offs in building a robust decision module, dissecting the critical systems properties of predictability and generalization. This culminates in a key insight into the conflict between benchmark alignment and real-world robustness.

    \item \textbf{A Unified, Generalizable Framework and its End-to-End Validation:} We formalize and implement DynaQuery, a unified framework integrating our robust linker with structured (SQP) and multimodal (MMP) pipelines. We then conduct a rigorous end-to-end validation of the framework's ability to generalize to a complex, "in-the-wild" database, identifying key architectural principles for building truly adaptable systems.
\end{itemize}

Through DynaQuery, we provide a validated blueprint for hybrid, self-adapting database interfaces, bridging the gap between high-level vision and practical application.
%================================================================================
%================================================================================

\section{Background and Conceptual Framework}
This section formally defines the core tasks addressed by our framework and situates our work within the context of relevant prior research. We first provide a formal problem formulation, then discuss the foundational systems and concepts that motivate our approach, and finally, we detail the specific technical challenge of schema linking.

\subsection{Formal Problem Formulation}
To establish the scope of our work, we formally define the two distinct yet related tasks that the DynaQuery framework is designed to solve.

\textbf{Database Definition:} Let a database $D$ be a collection of tables $T = \{t_1, t_2, \ldots, t_n\}$. Each table $t_i$ has a schema $S_i$ consisting of a set of columns $C_i = \{c_{i,1}, c_{i,2}, \ldots\}$, a set of primary keys $PK_i$, and a set of foreign keys $FK_i$. The full database schema is $S_D = \{S_1, S_2, ...\}$.

\textbf{Task 1: The NL-to-SQL Task.} Given a natural language query $Q_s$ and the database D, the goal is to generate a SQL query S such that its execution S(D) yields the correct answer set $R_s$ that satisfies the user's intent expressed in $Q_s$~\cite{yu18spider}.

\textbf{Task 2: The Multimodal Querying Task.} Given a natural language query $Q_m$ and the database D, where one or more columns $c_{i,j}$ may contain pointers to multimodal data $M$ (e.g., image URLs, document paths), the goal is to produce a filtered subset of records $R_m$ that satisfy semantic criteria expressed in $Q_m$, which require reasoning over the content of $M$ and are not answerable by standard SQL operations on $S_D$~\cite{chen23symphony, patel24lotus}.

\subsection{Foundational Systems and Concepts}
Our work is built upon and motivated by two key pillars of recent research: the grand vision for next-generation data systems articulated in the "Unbound Database" paper, and the conceptual framework for querying latent attributes established by the 'Generating Impossible Queries' (GIQ) system.

\textbf{The Unbound Database Vision.}
Our framework is designed as a direct and concrete instantiation of the "Unbound Database" vision proposed by Madden et al.~\cite{madden24unbound}. This vision calls for a new generation of data systems designed from the ground up to query the full spectrum of the world's data, from structured tables to unstructured assets. Table~\ref{tab:mapping} explicitly maps the core architectural concepts of this vision to their concrete implementations within DynaQuery, demonstrating how our work serves as a practical blueprint for this ambitious goal.

\begin{table*}[t]
\centering
\caption{Mapping Unbound Database Concepts ~\cite{madden24unbound} to DynaQuery's Concrete Components.}
\label{tab:mapping}
\begin{tabular}{lp{0.7\textwidth}}
\toprule
\textbf{Unbound Database Concept} & \textbf{DynaQuery Implementation} \\
\midrule
Declarative Interface & Natural Language query input, processed by the SILE as a "what, not how" request. \\
Logical Operators (e.g., Filter) & The core semantic reasoning step within the MMP, which acts as a powerful \textbf{semantic Filter} over records based on their linked unstructured content. \\
Multi-Objective Optimization & Our framework's modular design \textbf{enables} multi-objective trade-offs. The pluggable Decision Module (RQ2) demonstrates this by allowing a choice between a high-cost/high-generalization LLM and a low-cost/specialist BERT model. \\
Physical Plan Operators & The specific Chain-of-Thought prompts and classification modules used within our pipelines, which represent concrete implementations of logical reasoning steps. \\
Code Generation / Optimization & The Zero-Shot NL-to-SQL Pipeline (SQP), which generates efficient SQL code to delegate structured query processing to the database engine. \\
Extensibility ("Recipe Box") & The engineered prompts within SILE and the pipelines serve as reusable "recipes" for consistent reasoning. The modular architecture also allows for new pipelines to be added. \\
\bottomrule
\end{tabular}
\end{table*}
\textbf{The 'Generating Impossible Queries' (GIQ) System.}
The technical basis for our multimodal pipeline is the GIQ system~\cite{nejjar25giq}, which first established the use of multimodal LLMs to query latent, semantic attributes in databases. The work successfully demonstrated that an LLM could reason over linked text and image data to answer queries unsolvable by standard SQL. However, the GIQ implementation was a proof-of-concept on a controlled, single-table schema. It left the critical engineering challenges of generalization—such as schema-dependence, multi-table scope, and dynamic content discovery—unsolved. A primary contribution of DynaQuery is to address these challenges, thereby operationalizing the GIQ concept for complex, real-world databases.

\subsection{The Schema Linking Problem in NL-to-SQL}

Accurate schema linking—the task of correctly identifying the tables and columns in a database schema that are referenced in a natural language query—is a critical prerequisite for any Text-to-SQL system~\cite{maamari24death}. Foundational work by Lei et al.~\cite{lei20reexamining} provided the first systematic, data-driven study of this task, arguing that schema linking is not a minor component but the very "crux" of the Text-to-SQL problem.

In the modern context of LLMs, the challenge is no longer proving the importance of schema linking, but rather developing robust and scalable mechanisms to perform it. As noted in recent work, providing the full, unfiltered schema of a large database directly into a query-generation prompt can overwhelm an LLM's context window and degrade performance~\cite{gao24text2sql}. Therefore, an effective linking and pruning mechanism is essential. Approaches for this generally fall into two categories, which we compare in our experiments:

\textbf{Unstructured Context Retrieval (RAG):} This paradigm, introduced by Lewis et al.\cite{lewis20rag}, reframes the structured schema linking problem as an unstructured information retrieval task. While sophisticated systems like RASL\cite{eben24rasl} employ intelligent decomposition to avoid fracturing relational integrity, the retrieval-first paradigm remains architecturally flawed. The process is a decoupled, two-phase model: a probabilistic retrieval phase precedes a deterministic reasoning phase. Much like a query plan that prematurely discards data, if the initial retrieval—acting as a lossy filter—fails to surface a necessary table or transitive dependency, that structural context is irrevocably lost. The downstream generator is thus forced to reason over an incomplete schema, leading to incorrect join paths~\cite{chen-etal-2024-table} and fundamentally subordinating guaranteed structural integrity to the probabilistic recall of an information retrieval model.

\textbf{Structure-Aware Holistic Analysis:} In contrast to unstructured retrieval, our work adopts a structure-aware, holistic approach that aligns with state-of-the-art methods prioritizing relational integrity~\cite{pourreza24dinsql}. This two-stage process first leverages an LLM's reasoning capabilities over a complete, structured view of the schema to identify the relevant tables. In a second, programmatic pruning step, it uses this selection to construct a minimal schema context. This method is architecturally designed to preserve the global relational context, thereby avoiding the risk of irrecoverable context pruning introduced by RAG's probabilistic, retrieval-first architecture.

%================================================================================
\section{The DynaQuery Self-Adapting Framework}
The architecture of DynaQuery is founded on the philosophy of creating a modular, "plug-and-play" system that decouples query processing logic from the physical database schema. This modularity is key to our framework's flexibility and allows it to incorporate established LLM engineering patterns, such as Chain-of-Thought (CoT) prompting to elicit sophisticated reasoning~\cite{wei22cot}. Unlike static systems requiring manual configuration for each new database, DynaQuery is \textbf{self-adapting}. It uses a central schema introspection and linking engine to programmatically discover the schema and dynamically configure its query pipelines at runtime. This design aligns with the modular, lifecycle-based view of modern NL2SQL systems, as surveyed by Liu et al.~\cite{liu24survey}.

\subsection{Architectural Overview}
\begin{figure*}[t]
    \centering
    \includegraphics[width=\textwidth]{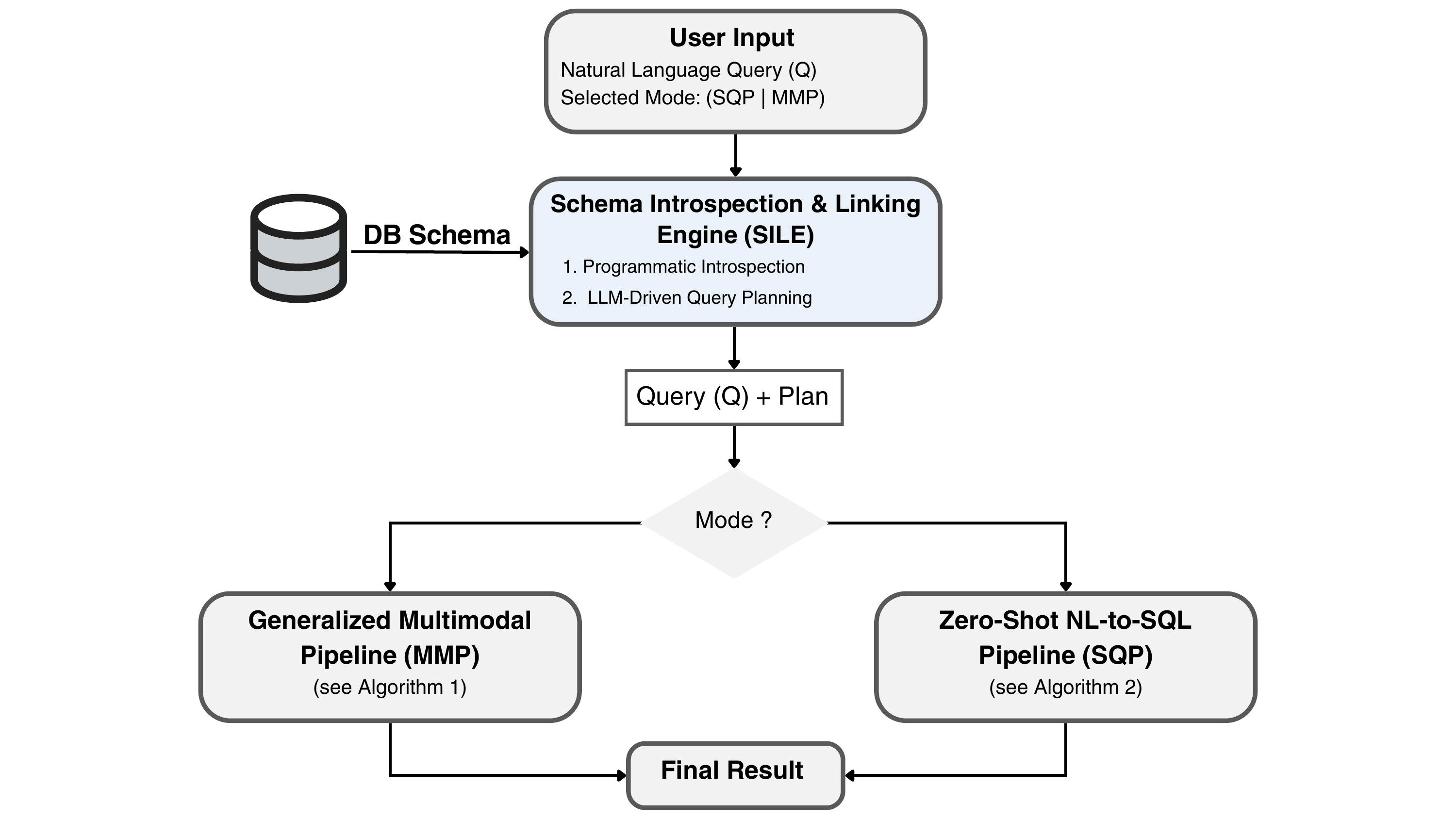}
    \caption{The high-level architecture of the DynaQuery framework}
    \label{fig:architecture}
\end{figure*}

As illustrated in Figure~\ref{fig:architecture}, the DynaQuery framework is composed of a central engine and two specialized pipelines. The process begins when a natural language query $Q$, first enters the \textbf{Schema Introspection and Linking Engine (SILE)}. The SILE acts as a universal pre-processor, analyzing the query against the database schema to produce a high-level query \textbf{plan}. 
This plan, which includes the relevant tables and their relationships, forms the foundation for the framework's two specialized pipelines. Our design prioritizes explicit user control. Rather than attempting automated task classification, the user directs the query and its generated plan to one of two specialized pipelines: the \textbf{Generalized Multimodal Pipeline (MMP)} for complex, semantic queries over unstructured data, or the \textbf{Zero-Shot NL-to-SQL Pipeline (SQP)} for structured data retrieval. This modular, orchestrated architecture is realized using the LangChain framework~\cite{chase22langchain}.

\subsection{The Core Enabler: The Schema Introspection and Linking Engine (SILE)}
The SILE is the cornerstone of the framework's adaptability, acting as a first-class query planning primitive. Its design combines programmatic schema discovery with LLM-driven strategic reasoning.

\textbf{Programmatic Introspection:}
Upon its first connection to a database, the SILE programmatically inspects the database's catalog to construct a detailed, in-memory representation of the full schema, $S_D$. This representation includes all tables, columns, data types, and relational constraints (e.g., primary and foreign keys). This structured metadata is generated once and cached, providing a comprehensive and efficient foundation for all subsequent query planning operations.

The SILE elevates schema linking from a simple selection task to a strategic query planning phase. The full schema representation is provided as context to an LLM, which is prompted to generate a structured query plan~\cite{maamari24death}. To enhance the robustness of this process, we leverage Chain-of-Thought (CoT) prompting~\cite{wei22cot}, encouraging the model to explicitly reason about the entities in the user's query and their mapping to the schema before producing the final plan. This plan is not merely a list of relevant tables; it is a conceptual blueprint that distinguishes between the \textit{base table} (containing the primary entity of the user's request) and any necessary \textit{join tables}. By producing a structured, reasoned plan, the SILE provides a robust strategic foundation for the downstream pipelines.

\subsection{Pipeline A: The Generalized Multimodal Pipeline (MMP)}
This pipeline is our primary contribution, designed to operationalize and generalize the conceptual framework of the GIQ system for complex, multi-table schemas. The architecture is founded on the principle of delegating structured filtering to the database engine to ensure scalability, a principle actively explored in modern query optimization~\cite{yang23predicate}, which we term the "Ultimate Filtered Join". The overall workflow is formalized in Algorithm~\ref{alg:mmp}, and its stages are detailed below.

\begin{algorithm}[htbp]
\caption{The Generalized Multimodal Pipeline (MMP) Workflow.}
\label{alg:mmp}
\begin{small} 
\begin{algorithmic}[1]
\Require User query $Q$, full schema $S_D$, database instance $D$
\Ensure Final set of accepted records $R_{\text{final}}$

\Function{GetCandidateSet}{$Q, S_D, D$} \Comment{Phase 1: Optimized Retrieval}
    \State $Plan \gets \textsc{SILE}(Q, S_D)$
    \State $S'_D \gets \textsc{PruneSchema}(S_D, Plan.\text{tables})$
    \State $C_{\text{multi}} \gets \textsc{DiscoverMultimodalColumns}(S'_D, D)$
    \State $C_{\text{where}} \gets \textsc{WhereClauseChain}(Q, S'_D, Plan)$
    \State $C_{\text{join}} \gets \textsc{JoinClauseChain}(S'_D, Plan)$
    \State $C_{\text{wnn}} \gets \textsc{BuildIsNotNullClause}(C_{\text{multi}})$
    \State $SQL_{\text{cand}} \gets \textsc{AssembleSQL}(\text{"SELECT *"}, C_{\text{join}}, C_{\text{where}}, C_{\text{wnn}})$
    \State $R_{\text{cand}} \gets \textsc{ExecuteSQL}(SQL_{\text{cand}}, D)$
    \State \Return $(R_{\text{cand}}, Plan, C_{\text{multi}})$
\EndFunction

\Function{FilterCandidates}{$Q, R_{\text{cand}}, C_{\text{multi}}$} \Comment{Phase 2: Multimodal Reasoning}
    \State $R_{\text{acc\_keys}} \gets \emptyset$
    \State $M_{\text{idx}} \gets \textsc{MapColsToIndices}(C_{\text{multi}}, R_{\text{cand}}.\text{cols})$
    \ForAll{record $r \in R_{\text{cand}}$}
        \State $Z_r \gets \textsc{GenerateRationale}(Q, r, M_{\text{idx}})$
        \State $label \gets \textsc{DecisionModule}(Q, Z_r)$
        \If{$label = \text{'ACCEPT'}$}
            \State Add $\text{PrimaryKey}(r)$ to $R_{\text{acc\_keys}}$
        \EndIf
    \EndFor
    \State \Return $R_{\text{acc\_keys}}$
\EndFunction

\Statex \textit{// Main Execution}
\State $(R_{\text{cand}}, Plan, C_{\text{multi}}) \gets \Call{GetCandidateSet}{Q, S_D, D}$
\State $R_{\text{acc\_keys}} \gets \Call{FilterCandidates}{Q, R_{\text{cand}}, C_{\text{multi}}}$
\If{$R_{\text{acc\_keys}}$ is not empty}
    \State $sql \gets \text{"SELECT * FROM "} + Plan.\text{base\_table}$
    \Statex \hspace{\algorithmicindent} $ + \text{" WHERE pk IN "} + R_{\text{acc\_keys}}$
    \State $R_{\text{final}} \gets \textsc{ExecuteSQL}(sql, D)$
\Else
    \State $R_{\text{final}} \gets \emptyset$
\EndIf
\State \Return $R_{\text{final}}$
\end{algorithmic}
\end{small}
\end{algorithm}
\begin{enumerate}
    \item \textbf{Query Planning and Schema Pruning (Lines 2--3):} The workflow begins by invoking the SILE to generate a high-level query plan. This plan identifies all tables required to answer the query, which are then used to create a minimal, pruned schema context, denoted as $S_D'$.

    \item \textbf{Optimized Query Assembly (Lines 4--8):} To construct a highly selective SQL query, the MMP orchestrates several specialized components. A key optimization is performed first: the system programmatically discovers which columns in the pruned schema contain multimodal data (\textbf{Line 4}). This allows the final query to be constructed to pre-filter any rows that lack the necessary unstructured content. Concurrently, LLM-powered chains generate the structured \texttt{WHERE} conditions (\textbf{Line 5}) and the necessary \texttt{JOIN} logic (\textbf{Line 6}). These components are then assembled into a single, optimized SQL query (\textbf{Line 8}).

    \item \textbf{Candidate Set Retrieval (Line 9):} The assembled SQL query is executed, retrieving a small and manageable set of candidate records, $R_{\text{cand}}$. This step leverages the database's query optimizer to perform the vast majority of the filtering work efficiently.

    \item \textbf{Per-Record Multimodal Reasoning (Lines 15--17):} Each record in the pre-filtered candidate set is then processed in a loop. A powerful multimodal LLM generates a rationale using Chain-of-Thought prompting~\cite{wei22cot}, evaluating the record's unstructured content against the user's semantic criteria. This rationale is then passed to a pluggable Decision Module to assign a final label.

    \item \textbf{Final Decision and Answer Synthesis (Lines 18--22, 26--31):} If a record is classified as 'ACCEPT', its primary key is collected. After all candidates are processed, these keys are used to execute a final, simple query to retrieve the full records, producing the answer set $R_{\text{final}}$. This two-phase process ensures that expensive multimodal reasoning is only performed on a small, highly-relevant, and pre-qualified subset of the data.
\end{enumerate}

\subsection{Pipeline B: The Zero-Shot NL-to-SQL Pipeline (SQP)}
This complementary pipeline is responsible for efficiently and accurately handling purely structured queries in a zero-shot manner~\cite{gao24text2sql}. Its straightforward workflow, formalized in Algorithm~\ref{alg:sqp}, leverages the SILE for robust context before delegating the core generation task to a powerful LLM.

\begin{algorithm}[htbp]
\caption{The Zero-Shot NL-to-SQL Pipeline (SQP) Workflow.}
\label{alg:sqp}
\begin{small} % Match the font size of Algorithm 1 for consistency
\begin{algorithmic}[1]
\Require User query $Q$, full schema $S_D$, database instance $D$
\Ensure Final SQL query result set $R_{\text{final}}$

\Statex \textit{// Phase 1: Contextualization}
\State $Plan \gets \text{SILE}(Q, S_D)$
\State $S'_D \gets \textsc{PruneSchema}(S_D, Plan.\text{tables})$

\Statex \textit{// Phase 2: Generation and Execution}
\State $SQL_{\text{raw}} \gets \textsc{GenerateSQLZeroShot}(Q, S'_D)$
\State $SQL_{\text{final}} \gets \textsc{SanitizeSQL}(SQL_{\text{raw}})$
\State $R_{\text{final}} \gets \textsc{ExecuteSQL}(SQL_{\text{final}}, D)$

\State \Return $R_{\text{final}}$
\end{algorithmic}
\end{small}
\end{algorithm}

The workflow proceeds in three stages:

\begin{enumerate}
    \item \textbf{Contextualization (Lines 1--2):} Similar to the MMP, the SQP's first step is to invoke the SILE to obtain a query plan and a pruned schema context, $S'_D$. Providing this minimal, high-fidelity context is crucial for preventing LLM confusion and improving the accuracy of the generated SQL~\cite{maamari24death}.

    \item \textbf{Query Generation (Line 3):} The linked schema and the user's query are passed to a powerful LLM using a carefully engineered zero-shot prompt. This prompt instructs the model to act as a SQL expert, leveraging its extensive pre-trained knowledge to generate a syntactically correct SQL query.

    \item \textbf{Sanitization and Execution (Lines 4--5):} The raw SQL output from the LLM is passed through a rule-based sanitization layer. This layer performs two critical functions. First, it cleans the LLM output by removing common artifacts (e.g., markdown code blocks). Second, and more importantly, it acts as a \textbf{safety guardrail}. The sanitizer programmatically ensures that only \texttt{SELECT} statements are extracted and passed to the database for execution, explicitly preventing any data modification (DML) or definition (DDL) operations~\cite{chafik25security}. The cleaned, validated query is then executed to produce the final result set.
\end{enumerate}
\section{Experimental Design}
The primary goal of our experimental evaluation is not merely to report isolated performance metrics, but to rigorously validate the central claims of this paper. Our experiments are designed to dissect our framework's architecture and validate its core components in a bottom-up fashion: we first validate our foundational linking primitive, then analyze a critical sub-component of our multimodal pipeline, and finally, test the end-to-end generalization of the unified system. This methodology is structured to directly answer the following research questions:

\subsection{Research Questions (RQs)}
\begin{itemize}
    \item \textbf{RQ1 (Foundational Component Validation):} Does the holistic, structure-aware analysis provided by our Schema Introspection and Linking Engine (SILE) yield higher accuracy on complex NL-to-SQL tasks compared to a standard unstructured RAG baseline?
    
    \item \textbf{RQ2 (Sub-Component Analysis):} In the context of our multimodal pipeline, how does a zero-shot, LLM-native classification module compare to a fine-tuned BERT baseline, and what are the architectural trade-offs regarding robustness and predictability?
    
    \item \textbf{RQ3 (End-to-End System Generalization):} Can our unified DynaQuery framework successfully generalize beyond standard benchmarks to operate effectively on a complex, "in-the-wild" database, handling a diverse spectrum of both structured and multimodal queries?
\end{itemize}

\subsection{Datasets}
To rigorously evaluate each component of our framework, we selected and constructed a diverse set of datasets tailored to each research question.

\textbf{For Schema Linking (RQ1):}
\begin{itemize}
    \item \textbf{Spider~\cite{yu18spider}:} Our primary evaluation for \textbf{RQ1} uses the widely-used Spider development set to test performance in a syntactically complex, cross-domain environment. For our direct linker evaluation, we use the accompanying ground-truth labels from the Spider-Schema-Linking Split~\cite{taniguchi21investigation}.
    \item \textbf{BIRD~\cite{li23bird}:} We use the BIRD development set to stress-test our linking architecture in a more complex, realistic setting that includes noisy data and external knowledge requirements. The rationale for selecting both Spider and BIRD is detailed in Section~\ref{sec:evaluation-protocol}.
\end{itemize}

\textbf{For Classifier Comparison (RQ2):}
\begin{itemize}
    \item \textbf{Annotated Rationale Dataset:} To provide a robust training and evaluation corpus for the decision module, we created and manually annotated a new dataset of 5,000 (Question, Rationale, Label) triplets. To ensure diversity, the rationales were generated using a variety of LLMs and prompt styles. Each triplet was then manually annotated by an author with a label from {ACCEPT, RECOMMEND, REJECT}. These labels correspond to whether the rationale provides a fully correct, partially correct, or incorrect justification for answering the question, respectively. The dataset is balanced across the three classes and will be made publicly available. For our experiments, we use a standard 80/20 split for training and in-distribution (IID) testing.
    \item \textbf{Out-of-Distribution (OOD) Case Study Set:} To evaluate true model robustness, we created a small, challenging OOD set from a novel e-commerce domain. This set consists of 45 `(Question, Rationale)` pairs, manually labeled according to a strict "hierarchical intent" philosophy where primary constraints are prioritized over secondary filters. This set is used for our qualitative case study.
\end{itemize}

\textbf{For Framework Generalization (RQ3):}
\begin{itemize}
\item \textbf{Olist Multimodal Benchmark:} To provide a rigorous end-to-end test for \textbf{RQ3}, we constructed a new benchmark based on the public Olist E-commerce dataset~\cite{olist_andr__sionek_2018}. We augmented this complex, multi-table relational schema by manually linking 100 products to high-quality images, creating a testbed that embodies the dual challenge of structured complexity, semantic ambiguity, and multimodal reasoning. For evaluation, we designed a suite of 40 novel questions. This suite is divided into two parallel sets of 20 queries each: one for the MMP (containing multimodal constraints) and one for the SQP (containing purely structured constraints). To ensure comprehensive evaluation, both 20-query sets were manually stratified by difficulty according to the official Spider hardness criteria (Easy, Medium, Hard, Extra Hard)~\cite{yu18spider}. The full benchmark, including our curation methodology and queries, is provided in our artifact repository.
\end{itemize}

\subsection{Baselines and Architectural Alternatives}
Our experiments are designed to compare our architectural choices against strong, relevant baselines and alternative configurations.

\begin{itemize}
    \item \textbf{For RQ1 (Linking Architectures):} To test our central hypothesis on schema linking, we compare our structure-aware SILE against a strong \textbf{unstructured RAG baseline}. Our RAG implementation is designed for a rigorous, head-to-head comparison of the core retrieval strategy. It follows established best practices for foundational RAG from recent surveys~\cite{wang24rag}, utilizing a state-of-the-art embedding model (BAAI/bge-large-en-v1.5)~\cite{xiao24cpack}. To isolate the effect of the initial context retrieval—the core of our hypothesis—we deliberately exclude orthogonal, post-retrieval optimizations like reranking, a common step in advanced RAG pipelines~\cite{wang24rag}, ensuring a fair and direct comparison.

    \item \textbf{For RQ2 (Decision Modules):} Our analysis of the decision module is an architectural trade-off study. We compare our proposed zero-shot generalist (an LLM-native classifier) against a strong specialist baseline: a \textbf{BERT-based classifier fine-tuned}~\cite{devlin19bert} on a large, in-domain dataset of 4,000 labeled examples.
\end{itemize}

\subsection{Evaluation Protocol for RQ1}\label{sec:evaluation-protocol}
To rigorously test our central hypothesis for RQ1—that a structure-aware linker is superior to an unstructured RAG baseline—we designed a multi-faceted evaluation across two key benchmarks that represent the evolution of the Text-to-SQL field.

\subsubsection{Benchmark Selection Rationale}
The choice of benchmarks is critical for a meaningful evaluation. We selected Spider and BIRD for the following strategic reasons:

\begin{itemize}
    \item \textbf{Spider~\cite{yu18spider}:} We include Spider as a foundational benchmark to validate our system's baseline proficiency in generating \textbf{syntactically complex SQL}. While acknowledging the well-documented risk that its contents may have contaminated the pre-training corpora of modern LLMs~\cite{ranaldi24contamination}, it remains an essential standard for comparing SQL generation logic. Crucially, the availability of the \textbf{Spider-Schema-Linking Split~\cite{taniguchi21investigation}} allows us to perform a complete, two-part analysis, providing a clear causal link between linking quality and final performance.

    \item \textbf{BIRD~\cite{li23bird}:} Our primary end-to-end evaluation is conducted on BIRD. Sourced from real industry data platforms, BIRD presents a more robust test of a system's practical utility by introducing challenges such as noisy database values, external knowledge requirements, and efficiency considerations. As it is a more recent and complex benchmark, it is also less susceptible to data contamination. We chose BIRD over the newer Spider 2.0 as our work focuses on the foundational challenge of single-shot query generation over complex \textit{data}, which is BIRD's core focus, rather than the multi-turn agentic \textit{workflows} introduced in Spider 2.0~\cite{lei24spider2}.
\end{itemize}

\subsubsection{Spider Evaluation Protocol: Random Sampling with Post-Hoc Validation}
Our evaluation on Spider is shaped by a core characteristic of the benchmark: its difficulty labels ('easy', 'medium', 'hard', 'extra') are an \textbf{implicit property}. They are not pre-assigned but can only be determined by parsing the gold SQL query and counting its syntactic components~\cite{yu18spider}.

\textbf{Sampling Strategy.} In the absence of pre-existing categories for stratification, we employed the standard and methodologically appropriate approach: random sampling. Our initial sample consisted of 500 queries from the annotated Spider development split~\cite{taniguchi21investigation}; a programmatic cross-referencing against the canonical Spider benchmark~\cite{yu18spider} data yielded a final, unambiguous set of 475 queries for all experiments.

\textbf{Execution Protocol.} To measure the semantic correctness of the generated SQL, we employ a direct execution protocol. As highlighted in multiple recent studies~\cite{pourreza23evaluating, zeng23reranking}, the official Spider evaluation script relies on a brittle internal parser that is sensitive to non-semantic stylistic variations. To ensure a more robust evaluation, our protocol bypasses this parsing logic entirely. A predicted query is deemed correct if and only if its execution against the database yields a result set identical to that of the ground-truth query, with appropriate handling for both ordered and unordered results. The specific metrics derived from this protocol are detailed in Section~\ref{sec:metrics}.

\textbf{Post-Hoc Analysis for Representativeness.} To scientifically validate our random sample, we performed a crucial post-hoc analysis. The official tool for classifying query difficulty is dependent on the aforementioned brittle parser. We therefore developed a robust, heuristic-based classifier that is directly modeled on the classification rules from the official evaluation script~\cite{yu18spider}. This approach provides a reliable and reproducible hardness classification without relying on the legacy parser. Our analysis using this classifier confirmed that the difficulty distribution of our 475-query sample is statistically indistinguishable from the full 1,034-query development set, validating that our results are representative and generalizable.
\subsubsection{BIRD Evaluation Protocol: Stratified Sampling with Explicit Labels}
Unlike Spider, the BIRD benchmark includes an explicit, human-annotated difficulty label ('simple', 'moderate', 'challenging') for every query~\cite{li23bird}. This key difference dictates a different sampling strategy and allows for the direct use of stratified random sampling.

\textbf{Sampling Strategy.} We performed stratified random sampling to construct a reproducible, 500-entry sample from the development set. This approach guarantees representativeness by design, deliberately creating a sample whose difficulty distribution is proportional to that of the full benchmark. This ensures our evaluation on BIRD is perfectly aligned with the challenges defined by the benchmark's creators.

\textbf{Metrics and Execution.} To measure both official metrics from the BIRD benchmark~\cite{li23bird}—Execution Accuracy (EA) and Valid Efficiency Score (VES)—while maintaining experimental efficiency, we adopted a multi-phase protocol. First, we generated prediction files for the full 1,534-entry development set, populating our 500 sampled queries and using placeholders for the rest to satisfy the official tooling requirements. Second, we ran minimally modified versions of the official evaluation tools to capture per-query correctness and performance. Finally, a post-processing script filtered these detailed reports to calculate the final EA and VES scores based only on our 500-query stratified sample. This protocol ensures our results are both reproducible and directly comparable to the broader field.
\subsection{Metrics}\label{sec:metrics}
Two standard metrics are used to evaluate Text-to-SQL systems: Exact Match (EM) and Execution Accuracy (EA)~\cite{luo25survey, yu18spider}. While EM evaluates syntactic form, EA measures true semantic correctness by comparing the final result sets. As recent studies have extensively documented the unreliability of EM~\cite{pourreza23evaluating}, the field has largely adopted Execution Accuracy as the primary metric for correctness. For classification sub-tasks, we use the standard metrics of Precision (\(P\)), Recall (\(R\)), and F1-Score, defined for a given class \(c\) as:
\begin{align*}
P_c &= \frac{TP_c}{TP_c + FP_c} \\
R_c &= \frac{TP_c}{TP_c + FN_c} \\
F1_c &= 2 \cdot \frac{P_c \cdot R_c}{P_c + R_c}
\end{align*}
Our evaluation focuses on the following specific metrics for each research question:
\begin{itemize}
    \item \textbf{Metrics for RQ1 (Linking Architectures):} Our evaluation of linking architectures is two-fold:
    \begin{itemize}
        \item \textit{Direct Linking Performance:} For the isolated sub-task of identifying the correct tables, we measure standard Precision, Recall, and F1-Score.
        \item \textit{End-to-End Correctness:} For the full Text-to-SQL task, our primary metric is \textbf{Execution Accuracy (EA)}, which measures the fraction of generated queries that produce the correct result set when executed. It is formally defined as:
        \[
        \text{EA} = \frac{1}{|Q|} \sum_{q \in Q} \mathbb{I}(\text{exec}(S_{pred}) \equiv \text{exec}(S_{gold}))
        \]
        where $\mathbb{I}(\cdot)$ is the indicator function for correctness.
        \item \textit{End-to-End Efficiency (BIRD only):} For our evaluation on the BIRD benchmark, we also report the \textbf{Valid Efficiency Score (VES)}~\cite{li23bird}. VES combines correctness with performance by measuring the runtime of valid queries relative to the human-written gold standard. It is formally expressed as:
        \[
        \text{VES} = \frac{1}{|Q|} \sum_{q \in Q} \mathbb{I}(\text{EA}) \cdot \sqrt{\frac{\text{time}(S_{gold})}{\text{time}(S_{pred})}}
        \]
    \end{itemize}

    \item \textbf{Metrics for RQ2 (Classifier Performance):} For our three-class classification task, we report macro-averaged Precision, Recall, and F1-Score.
\end{itemize}

\subsection{Implementation Details}

\textbf{LLM Configuration.} All Large Language Model components in our framework, including the SILE, the SQL generation chains (for both SQP and MMP), and the LLM-native Decision Module, were implemented using the Google Gemini API. We utilized the \texttt{gemini-2.5-pro} (Stable release, June 17, 2025) model version~\cite{google25gemini}. To ensure deterministic and reproducible outputs for our experiments, all API calls were made with a temperature of 0.0.

\textbf{Fine-Tuned Classifier.} The fine-tuned specialist baseline for our RQ2 evaluation was based on the \texttt{bert-base-cased} model from the Hugging Face Transformers library~\cite{wolf20transformers}. The model was fine-tuned for 4 epochs on our \textit{Annotated Rationale Dataset} using the AdamW optimizer~\cite{loshchilov19adamw}, with a batch size of 32 and a learning rate of 2e-5.

\textbf{Hardware and Frameworks.} All experiments were conducted on a workstation equipped with a single NVIDIA RTX 4090 GPU (24GB VRAM). The DynaQuery framework is implemented in Python 3.10, and the LLM orchestration is built upon the LangChain library~\cite{chase22langchain}. The database backend used for all experiments was MySQL version 9.3.0.
\section{Results and Analysis}
This section presents the results of our experiments, organized by our three research questions. We first validate our foundational SILE component (RQ1), then perform a deep-dive into a key design choice within the MMP (RQ2), and finally, demonstrate the end-to-end generalization of the full DynaQuery framework on a challenging, real-world benchmark (RQ3).

\subsection{RQ1: The Primacy of Structure-Aware Linking}
To answer RQ1, we evaluated our core SILE component against a strong RAG baseline across the Spider and BIRD benchmarks. This experiment was designed to test our central hypothesis: that a holistic, structure-aware approach to schema linking is architecturally superior to treating the schema as an unstructured document for retrieval.

\subsubsection{Part 1: Component-Level Linking Performance}
We first isolated the linking components and measured their ability to identify the correct set of tables on our 475-query sample of the Spider development set. The results, presented in Table~\ref{tab:linking_perf}, reveal a categorical difference in performance.

\begin{table}[h]
\centering % Added for better centering
\caption{Direct Schema Linking Performance on the Spider-dev Set (475-entry sample). Higher is better.}
\label{tab:linking_perf}
\begin{tabular}{lccc}
\toprule
\textbf{Method} & \textbf{Precision} & \textbf{Recall} & \textbf{F1-Score} \\
\midrule
RAG Baseline & 23.6\% & 64.0\% & 32.9\% \\
\textbf{DynaQuery (SILE)} & \textbf{73.0\%} & \textbf{85.6\%} & \textbf{77.0\%} \\
\bottomrule
\end{tabular}
\end{table}

The underlying metrics expose the fundamental architectural weakness of the RAG approach for this task. While RAG achieves a reasonable Recall (64.0\%), its Precision is exceptionally low (23.6\%). This demonstrates that the retriever acts as a 'dragnet,' retrieving a noisy and confusing context bloated with irrelevant schema information. In contrast, our SILE's holistic analysis is vastly more precise (73.0\%). It acts as a 'scalpel,' preserving relational integrity to provide a clean, minimal, and structurally coherent context for the downstream query generator.

\subsubsection{Part 2: End-to-End Performance on Spider}
Next, we evaluated the end-to-end impact of these linkers on the final SQL generation task on Spider. The results in Table~\ref{tab:ea_breakdown} confirm that the superior context from SILE translates directly into a massive improvement in Execution Accuracy.

\begin{table}[h]
\centering
\caption{End-to-End Execution Accuracy (EA) on Spider by Query Difficulty. Delta indicates the absolute percentage point improvement.}
\label{tab:ea_breakdown}
\resizebox{\linewidth}{!}{%
\begin{tabular}{lccc}
\toprule
\textbf{Difficulty} & \textbf{RAG Baseline} & \textbf{DynaQuery (SILE)} & \textbf{Delta (Pts.)} \\
\midrule
Easy & 61.9\% & 91.8\% & \textbf{+29.9} \\
Medium & 55.9\% & 77.1\% & \textbf{+21.2} \\
Hard & 54.6\% & 75.3\% & \textbf{+20.7} \\
Extra & 61.8\% & 85.3\% & \textbf{+23.5} \\
\midrule
\textbf{Overall} & \textbf{57.1\%} & \textbf{80.0\%} & \textbf{+22.9} \\
\bottomrule
\end{tabular}%
}
\end{table}
The superior linking quality of the SILE translates directly into a massive performance gain in the end-to-end task. 
Overall, DynaQuery achieves an Execution Accuracy of 80.0\%, a \textbf{+22.9 absolute point improvement} over the RAG baseline. 
The nearly \textbf{30-point} delta on 'easy' queries is particularly revealing: it demonstrates that the noisy context from the RAG baseline is a major source of error even for structurally simple questions. 
This result confirms that in a syntactically complex environment like Spider, a structure-aware linking primitive is a critical prerequisite for high-fidelity query generation.
\subsubsection{Part 3: Stress-Testing on the BIRD Benchmark}
To validate the robustness of our architecture in a more realistic, data-grounded environment, we conducted our end-to-end evaluation on the BIRD benchmark. The results, presented in Table~\ref{tab:bird_ea_breakdown}, not only confirm the superiority of our structure-aware approach but demonstrate that its advantage widens significantly when faced with real-world data complexity.

\begin{table}[h]
\centering
\caption{End-to-End Execution Accuracy (EA) on BIRD by Query Difficulty (500-entry stratified sample).}
\label{tab:bird_ea_breakdown}
\begin{tabular}{lccc}
\toprule
\textbf{Difficulty} & \textbf{RAG (EA)} & \textbf{DynaQuery (EA)} & \textbf{Delta (Pts.)} \\
\midrule
Simple & 37.42\% & 66.23\% & \textbf{+28.81} \\
Moderate & 20.53\% & 45.03\% & \textbf{+24.50} \\
Challenging & 36.17\% & 53.19\% & \textbf{+17.02} \\
\midrule
\textbf{Overall} & \textbf{32.20\%} & \textbf{58.60\%} & \textbf{+26.40} \\
\bottomrule
\end{tabular}
\end{table}
This stress-test on BIRD reveals the key finding of our evaluation. While both systems' performance decreases on this more challenging benchmark, the \textit{manner} of their degradation exposes a fundamental architectural difference. The RAG baseline's performance \textbf{collapses}, plummeting from 57.1\% on Spider to just 32.2\%. In contrast, DynaQuery's accuracy \textbf{degrades gracefully}, with its lead over the baseline widening to a massive \textbf{+26.4 absolute points}. This divergence in performance under pressure provides strong evidence that unstructured retrieval is an unreliable foundation for this task. Our structure-aware SILE, however, proves to be a resilient architectural primitive, essential for building systems that can withstand the challenges of real-world data.
\subsubsection{Part 4: Programmatic Failure Analysis on BIRD}
To understand the root cause of this performance divergence, we developed a programmatic analysis pipeline using the \texttt{sqlglot} library~\cite{mao20sqlglot} to parse each failed query into its Abstract Syntax Tree (AST) for systematic categorization. The results, shown in Table~\ref{tab:failure_analysis}, reveal a pronounced, architectural contrast in the failure modes of the two systems.

\begin{table}[h]
\centering
\caption{Programmatic Failure Analysis on the BIRD Benchmark. Percentages are of total failures for each model.}
\label{tab:failure_analysis}
\begin{tabular}{lcc}
\toprule
\textbf{Error Category} & \textbf{RAG} & \textbf{DynaQuery} \\
\midrule
\textit{Contextual Failures} & \multicolumn{2}{c}{} \\
\quad \texttt{SCHEMA\_HALLUCINATION} & \textbf{50.74\%} & 6.76\% \\
\quad \texttt{JOIN\_TABLE\_MISMATCH} & 23.30\% & 26.57\% \\
\midrule
\textit{Logical \& Syntactic Failures} & \multicolumn{2}{c}{} \\
\quad \texttt{SELECT\_COLUMN\_MISMATCH} & 10.62\% & \textbf{33.82\%} \\
\quad \texttt{WHERE\_OR\_LOGIC\_ERROR} & 4.13\% & \textbf{19.32\%} \\
\quad Other Minor Errors & 11.21\% & 13.53\% \\
\bottomrule
\end{tabular}
\end{table}

The analysis provides quantitative evidence for the architectural brittleness of the RAG-based approach. Its failures are predominantly foundational, with over 50\% of its errors falling into the \texttt{SCHEMA\_HALLUCINATION} category: a failure mode where the model generates queries referencing non-existent tables or columns~\cite{kothyari23crush4sql}. This demonstrates that the unstructured retrieval mechanism frequently fails to provide a coherent and grounded context, causing the downstream LLM to detach from the reality of the database.

In stark contrast, DynaQuery almost entirely eliminates this catastrophic failure mode. For our system, the bottleneck shifts from contextual grounding to logical precision, with failures now concentrated in downstream reasoning tasks like identifying the correct select columns \texttt{SELECT\_COLUMN\_MISMATCH}. However, this analysis also reveals a crucial limitation of purely \textbf{schema-aware} linking. Our SILE excels at this task, correctly mapping query entities to schema names. The failures on BIRD, however, highlight a distinct and more challenging task: \textbf{data-aware linking}~\cite{li23bird}. This requires the system to correctly map a \textit{data value} mentioned in the query to its corresponding column in the schema, a task that often requires disambiguating between multiple potential columns across different tables. Since this value-to-column mapping cannot be resolved using schema metadata alone, our analysis concludes that while a structure-aware foundation is critical, the path to solving the full complexity of real-world queries requires this next level of data-aware reasoning.

\subsubsection{Part 5: Validating Query Quality with Efficiency Score (VES)}
To move beyond a binary measure of correctness and assess the practical utility of our systems, we evaluate the quality of the generated SQL using the Valid Efficiency Score (VES), a metric introduced by the BIRD benchmark specifically to address SQL performance on large-scale databases~\cite{li23bird}. VES elegantly combines correctness and performance, awarding a score of 0 to incorrect queries and a performance-based score to correct ones. Our results, presented in Table~\ref{tab:bird_ves_breakdown}, provide compelling evidence for the architectural superiority of our structure-aware approach in generating not just correct, but also high-quality, efficient SQL.

\begin{table}[h]
\centering
\caption{Valid Efficiency Score (VES) on the BIRD Benchmark. The performance gain of DynaQuery (SILE) over the RAG baseline is shown in absolute points. Higher is better.}
\label{tab:bird_ves_breakdown}
\begin{tabular}{lccc}
\toprule
\textbf{Difficulty} & \textbf{RAG (VES)} & \textbf{DynaQuery (VES)} & \textbf{Gain (pts.)} \\
\midrule
Simple & 38.63 & 67.81 & \textbf{+29.18} \\
Moderate & 21.81 & 50.74 & \textbf{+28.93} \\
Challenging & 39.14 & 62.96 & \textbf{+23.82} \\
\midrule
\textbf{Overall} & \textbf{33.60} & \textbf{62.20} & \textbf{+28.60} \\
\bottomrule
\end{tabular}
\end{table}

Evaluating query efficiency reveals a dramatic performance gap. Overall, DynaQuery (62.20) achieves a score that is nearly double that of the RAG baseline (33.60), a massive performance gain of +28.6 points. This demonstrates that the clean, minimal, and relationally coherent context provided by SILE enables the downstream LLM to generate more optimal query plans. The low VES of the RAG baseline is a direct consequence of its low Execution Accuracy; since incorrect queries receive a VES of 0, its score is heavily penalized.

Notably, DynaQuery's overall VES (62.20) is higher than its Execution Accuracy (58.60\%). This indicates that for the queries it answers correctly, it frequently generates SQL that is more performant than the human-written gold standard, a phenomenon that aligns with recent studies questioning the optimality of benchmark ground-truth queries~\cite{pourreza23evaluating}. This phenomenon can be attributed to the "SILE Effect": by providing a simplified problem space, SILE allows the generator LLM to leverage its vast pre-trained knowledge of diverse SQL patterns to identify more direct relational paths. This often results in avoiding the unnecessary JOINs or inefficient subqueries that can arise from the noisy, bloated context provided by an unstructured RAG retriever. This finding confirms that a structure-aware linking primitive like SILE is a critical component for building Text-to-SQL systems that are not just functionally correct but also efficient enough for practical, real-world deployment.

\subsection{RQ2: Architectural Trade-offs in the Decision Module}
Having established the superiority of our structure-aware linking architecture in RQ1, we now turn our focus inward to a critical component of the Generalized Multimodal Pipeline (MMP): the final decision module. This module is responsible for the crucial last step of translating an LLM's unstructured, natural language rationale into a structured, discrete classification (\{ACCEPT, RECOMMEND, REJECT\}). The choice of architecture for this component is non-trivial and has significant implications for the overall system's robustness and predictability.

Our RQ2 evaluation therefore dissects the architectural trade-offs between two primary approaches: a \textbf{fine-tuned specialist} (BERT)~\cite{devlin19bert} and a \textbf{prompt-guided generalist} (LLM). Our analysis reveals that while the specialist excels on in-distribution data, it proves to be unpredictable and error-prone when faced with novel semantic challenges. In contrast, we show that a generalist LLM, when engineered with a precise, rule-based prompt, can be configured as a robust and predictable logical engine—a critical property for any reliable database system.
\subsubsection{Part 1: In-Distribution Performance and Prompt Alignment}
We first established a baseline by evaluating three classifier architectures on our 1,000-sample in-distribution (IID) test set. This dataset was annotated by a human expert with a nuanced, context-aware philosophy. The architectures are:

\begin{itemize}
\item \textbf{Fine-Tuned BERT:} A specialist model trained on 4,000 labeled examples~\cite{devlin19bert}.
\item \textbf{LLM (Descriptive Prompt):} A generalist LLM guided by a detailed, human-like prompt designed to mirror the annotation philosophy.
\item \textbf{LLM (Rule-Based Prompt):} A generalist LLM guided by a strict, logical prompt. This prompt instructs the model to follow a simple, literal interpretation of the labels by counting the user's constraints: ACCEPT if ALL constraints are met, RECOMMEND if SOME (but not all) are met, and REJECT if NONE are met.
\end{itemize}

\begin{table}[h]
\centering
\caption{Macro F1-Scores on the In-Distribution (IID) Test Set. The results highlight the effect of prompt-data alignment.}
\label{tab:rq2_iid_results}
\begin{tabular}{lc}
\toprule
\textbf{Classifier Architecture} & \textbf{Macro F1-Score} \\
\midrule
Fine-Tuned BERT (Specialist) & \textbf{99.1\%} \\
LLM (Descriptive Prompt) & 94.7\% \\
LLM (Rule-Based Prompt) & 78.0\% \\
\bottomrule
\end{tabular}
\end{table}

As shown in Table~\ref{tab:rq2_iid_results}, both the BERT model and the Descriptive Prompt LLM achieved high performance, as they are philosophically aligned with the test set's labels. Counter-intuitively, the logically stricter Rule-Based prompt performed worse. A manual analysis confirmed this was not due to model error, but a philosophical disagreement: the prompt's rigid ALL/SOME/NONE logic conflicted with the more nuanced, hierarchical judgments in the ground truth. This demonstrates that a classifier's performance on a static benchmark is not an absolute measure of its capability, but a measure of its alignment with the benchmark's labeling philosophy.

\subsubsection{Part 2: Out-of-Distribution (OOD) Qualitative Case Study}
To test for true generalization and robustness, we conducted a qualitative case study on a small, out-of-domain dataset. We designed a suite of three strategically chosen queries to stress-test different reasoning capabilities: simple visual matching, logical negation, and complex compound logic.

\begin{itemize}
    \item \textbf{Q1 (Simple Visual):} ``Show me products with green lid.''
    \item \textbf{Q2 (Negation \& Context):} ``Show me products that do not have the word 'baby' in their packaging image.''
    \item \textbf{Q3 (Compound Logic):} ``Show me products with a dispenser pump head, a rating $\geq 4.5$, and number of ratings $> 100.$''
\end{itemize}

We evaluated all three classifier architectures against a manually curated ground truth reflecting a nuanced, ``hierarchical intent'' philosophy. The results, summarized in Table~\ref{tab:rq2_ood_summary}, reveal the specialist model's failure to generalize and highlight a fundamental divergence in the behavior of the two generalist LLM configurations.

\begin{table}[h]
\centering
\caption{Accuracy of each classifier on the three OOD queries against our human-centric, hierarchical ground truth.}
\label{tab:rq2_ood_summary}
\begin{tabular}{lccc}
\toprule
\textbf{Query Type} & \textbf{BERT} & \textbf{LLM (Desc.)} & \textbf{LLM (Rule)} \\
\midrule
Q1: Simple Visual & 93.3\% & 100\% & 100\% \\
Q2: Negation & 73.3\% & 100\% & 100\% \\
Q3: Compound Logic & 73.3\% & 93.3\% & \textbf{20.0\%} \\
\midrule
\textbf{Overall Accuracy} & 78.9\% & 97.8\% & \textbf{73.3\%} \\
\bottomrule
\end{tabular}
\end{table}

Our qualitative analysis revealed three distinct and systematic behavioral profiles corresponding to the different classifier architectures.

\textbf{The Specialist: The Limits of Specialization.}
The fine-tuned BERT model, which demonstrated near-perfect accuracy on in-distribution data, struggled to generalize when faced with novel semantic structures. Its primary failure mode was an over-reliance on statistical patterns learned during fine-tuning. This was most evident in the logical negation task (Q2), where strong keyword associations (e.g., with the word "baby") appeared to override the explicit syntactic cues for negation within the rationale. Its inconsistent performance on the compound query (Q3) further suggests that its specialized knowledge did not equip it to handle complex, unseen logical compositions. This behavior is characteristic of specialist models: they achieve high performance by optimizing for known patterns but can fail unpredictably when encountering out-of-distribution phenomena for which no pattern has been learned.

\textbf{The Descriptive Generalist: The Challenge of Implicit Hierarchies.}
The LLM guided by a descriptive, human-like prompt successfully handled simple matching and negation. However, when faced with compound logic (Q3), it exposed a fundamental challenge in translating nuanced human intent into strict query semantics. The model struggled to infer the implicit priority of constraints, often treating them as a soft checklist. For instance, it would classify a record as RECOMMEND for satisfying secondary numerical criteria, even when it failed the primary structural constraint (e.g., lacking a "dispenser pump head"). While this behavior mimics a flexible, human-like interpretation, it conflicts with the precise, non-negotiable semantics required by a reliable data retrieval system.

\textbf{The Rule-Based Generalist: Predictable and Logically Consistent.}
Conversely, the Rule-based LLM demonstrated the properties of a truly robust and predictable system. Its strict, logical framework enabled it to flawlessly handle simple matching and negation. Its dramatically low accuracy on the compound query (Q3) is not a failure, but rather the key finding of this analysis. The model correctly executes a literal, boolean interpretation of the query, treating all constraints as equally important. For example, it correctly labels a product that fails the critical 'dispenser pump' constraint but meets a minor secondary one as RECOMMEND (since 1 of 3 conditions were met). While this diverges from our nuanced, hierarchical ground truth (which would label it REJECT), this \textbf{logical consistency} is a crucial and desirable property for a predictable database system. This fundamental trade-off between inferred intent and literal execution is analyzed further in our Discussion.

\begin{figure*}[t]
    \centering
    \includegraphics[width=\textwidth]{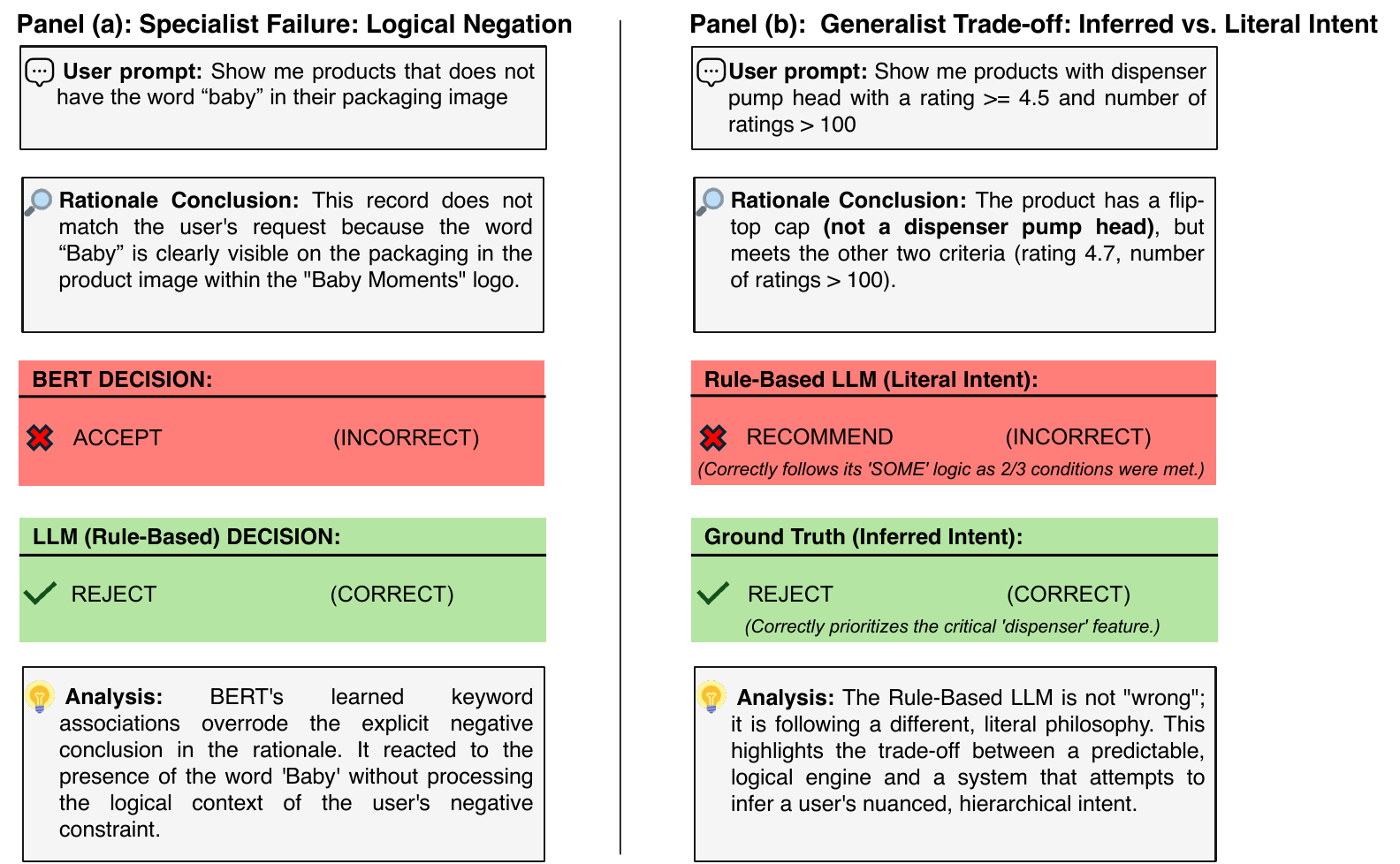}
    \caption{Qualitative analysis of systematic behaviors on Out-of-Distribution (OOD) queries. 
    Panel (a) shows the specialist's catastrophic failure on logical negation. 
    Panel (b) illustrates the fundamental conflict between a system that infers hierarchical intent versus one that executes literal, logical intent.}
    \label{fig:rq2_failures}
\end{figure*}

Figure~\ref{fig:rq2_failures} presents a side-by-side analysis of two "golden examples" from our case study. Panel (a) shows a catastrophic failure of the BERT specialist on a logical negation query (Q2). Panel (b) provides a clear illustration of the "Inferred vs. Literal Intent" conflict, showing how the Rule-Based LLM's logically consistent decision diverges from our human-centric ground truth on a complex compound query (Q3). These examples provide definitive, qualitative evidence for the superior robustness and predictability of the Rule-Based LLM architecture.

\subsection{RQ3: Validating Generalization of DynaQuery}
Our final and most challenging research question tests the end-to-end generalization of the full DynaQuery framework. To do this, we use our novel Olist benchmark, which was constructed to embody the dual challenges of \textbf{structured complexity and multimodal reasoning} while also introducing a high degree of real-world \textbf{semantic ambiguity}. Our evaluation proceeds in two stages. First, we measure the performance of our baseline, purely schema-aware framework to establish a performance baseline on this new, challenging task. Second, we introduce a minimal architectural adaptation, Schema Enrichment, and measure its impact on performance to test our hypothesis about the importance of semantics-awareness for generalization.
\subsubsection{Performance of the Baseline Schema-Aware Framework}
We first deployed the framework using the baseline, schema-aware SILE that proved effective in RQ1. We evaluated both pipelines on our 20-query, hardness-stratified Olist benchmarks. As shown in Table~\ref{tab:baseline_olist_results}, the performance of this baseline configuration was erratic and clearly limited by the benchmark's semantic complexity.

\begin{table}[htbp]
\centering
\caption{Performance of the baseline (schema-aware) DynaQuery framework on the Olist benchmark, by hardness.}
\label{tab:baseline_olist_results}
\begin{tabular}{lcc}
\toprule
\textbf{Hardness} & \textbf{MMP (F1-Score)} & \textbf{SQP (Exec.\ Acc.)} \\
\midrule
Easy & 57.77\% & 100.00\% \\
Medium & 60.00\% & 40.00\% \\
Hard & 59.00\% & 80.00\% \\
Extra Hard & 40.00\% & 40.00\% \\
\midrule
\textbf{Overall} & \textbf{54.20\%} & \textbf{65.00\%} \\
\bottomrule
\end{tabular}
\end{table}

The baseline framework's performance profile was inconsistent. While the SQP achieved perfect accuracy on simple queries, its performance was unpredictable on more complex tasks, dropping to 40.00\% on medium-difficulty queries before recovering to 80.00\% on hard queries. This erratic behavior, along with the MMP's generally low performance, pointed to a systematic issue. A failure analysis confirmed the root cause: the schema-aware SILE consistently failed to resolve the high degree of semantic ambiguity in the Olist schema (e.g., its implicit multilingual relationships). This result empirically demonstrates that pure schema-awareness, while a strong foundation, is insufficient for robust generalization to complex, real-world database schemas.

\subsubsection{Adaptation to Semantic Awareness and Final Performance}
This finding points towards the need for a more advanced linking capability. While our analysis in RQ1 suggests that a fully \textit{data-aware} linker is the ultimate goal, we hypothesize that a practical and powerful intermediate step is to make the system \textbf{semantics-aware}. To test this, we introduced a minimal architectural adaptation: \textbf{Schema Enrichment}. This technique enhances the SILE by allowing it to ingest an optional Semantic Schema Description file, providing human-readable comments for schema elements. This approach respects our “plug-and-play” philosophy, as providing a data dictionary is a cornerstone of good database management and a critical artifact for developers~\cite{linares16documenting}.

With this simple enhancement, we re-ran the benchmarks. The impact was transformative, as shown in Table~\ref{tab:final_framework_results}.

\begin{table}[htbp]
\centering
\caption{Final performance of the semantics-aware DynaQuery framework on the Olist benchmark by hardness.}
\label{tab:final_framework_results}
\begin{tabular}{lcc}
\toprule
\textbf{Hardness} & \textbf{MMP (F1-Score)} & \textbf{SQP (Exec.\ Acc.)} \\
\midrule
Easy        & 97.77\% & 100.00\% \\
Medium      & 95.00\% & 100.00\% \\
Hard        & 80.00\% & 100.00\% \\
Extra Hard  & 100.00\% & 80.00\% \\
\midrule
\textbf{Overall} & \textbf{93.20\%} & \textbf{95.00\%} \\
\bottomrule
\end{tabular}
\end{table}

The impact of this adaptation was transformative. With semantic context, the SQP's overall Execution Accuracy surged by 30 percentage points to a near-perfect \textbf{95.00\%}. The effect on the MMP was even more pronounced, with its F1-score increasing dramatically from 54.00\% to  \textbf{93.20\%}. This dramatic improvement validates our hypothesis that for complex, real-world schemas, bridging the 'semantic gap' is a critical prerequisite for generalization. Our Schema Enrichment technique provides a practical and effective mechanism to achieve this.

\subsubsection{Conclusion for RQ3}
This experiment validates that our full DynaQuery framework can successfully generalize to complex, "in-the-wild" databases, provided it is equipped with the necessary semantic context. It provides a key finding: for many ambiguous, real-world schemas, the critical step to unlock robust performance is to move from schema-awareness to \textbf{semantics-awareness}. Our Schema Enrichment technique provides a practical, low-overhead path to achieving this, representing a significant step towards truly adaptable natural language interfaces.
\section{Discussion}
Our experimental results do more than just validate our architectural choices; they provide novel insights into the fundamental challenges of building robust, generalizable natural language interfaces for databases. In this section, we synthesize our findings into three key discussions. First, we propose a conceptual framework, the "Hierarchy of Awareness," to organize the capabilities required for a system to handle diverse databases. Second, we analyze the inherent challenge of ensuring reasoning consistency in LLM-driven planning. Finally, we discuss the critical systems properties of predictability and controllability that emerged from our analysis of the decision module.
\subsection{The Hierarchy of Awareness: Schema, Semantics, and Data}
Our work empirically demonstrates a clear progression of capabilities necessary for a system to robustly handle diverse databases, a framework we term the Hierarchy of Awareness.

\textbf{Level 1: Schema-Awareness.} Our RQ1 results provide definitive, multi-benchmark evidence for the primacy of structured understanding. The catastrophic failure rate of the RAG baseline due to \texttt{SCHEMA\_HALLUCINATION}~\cite{kothyari23crush4sql} confirms that treating a database schema as an unstructured collection of text is an unreliable foundation. In contrast, our SILE’s programmatic introspection provides a grounded, high-fidelity context that nearly eliminates this failure mode. This establishes that a \textbf{schema-aware} architecture is the critical first step for any dependable Text-to-SQL system.

\textbf{Level 2: Semantics-Awareness.} While effective on well-struc\-tured benchmarks like Spider, the performance of our purely schema-aware system on the more complex BIRD benchmark hinted at the limits of this approach. Our RQ3 evaluation on the Olist database confirmed this hypothesis decisively. The Olist schema's high degree of semantic ambiguity, particularly its implicit cross-lingual relationships, caused the baseline SILE's performance to collapse. This led to our key architectural finding: to generalize to bespoke, real-world databases, a system must be enhanced to become \textbf{semantics-aware}. Our Schema Enrichment technique provides a practical, low-overhead mechanism to achieve this, and the dramatic performance recovery validates that bridging this 'semantic gap' is a necessary step for building truly adaptable systems.

\textbf{Level 3: Data-Awareness (The Next Frontier).} Finally, our failure analysis on BIRD points to the ultimate frontier. The SILE's occasional struggles with queries requiring value-based linking show that the final level of mastery is to become \textbf{data-aware}. This requires the system to correctly map a data value mentioned in the query to its corresponding column in the schema, a task that often requires disambiguating between multiple potential columns. Conquering this value-to-column mapping challenge, a core focus of modern benchmarks like BIRD~\cite{li23bird}, is the next critical step for the field.
\subsection{Reasoning Consistency as a Core Challenge}
A key finding from our RQ3 benchmark evaluation is the challenge of ensuring \textbf{reasoning consistency} in the schema linking process. Our framework demonstrated a remarkable capability to handle high structural complexity, successfully generating complex, multi-hop joins involving up to six tables for several 'Extra Hard' queries. Paradoxically, the system exhibited failures on structurally simpler queries that required a subset of the same relational paths, failing to include a necessary intermediate 'bridge' table in its query plan.
This paradoxical result—succeeding on a harder task while failing on a similar, easier one—is not a flaw in our system's architecture but rather exposes the inherent \textbf{unpredictability} of the LLM's zero-shot reasoning. This demonstrates that for Text-to-SQL systems, raw capability is not the only metric of success; \textbf{reliability and consistency are paramount}. Achieving deterministic, correct reasoning across all variations of natural language queries remains a critical open research problem, suggesting that future work must focus on making reasoning processes more robust and predictable through techniques like self-correction, a method with both practical agentic implementations~\cite{shinn23reflexion} and a growing theoretical foundation~\cite{wang24selfcorrection}.
\subsection{Predictability and Controllability as First-Class System Properties}
Our analysis of the MMP's Decision Module in RQ2 revealed a fundamental trade-off in natural language interfaces: the conflict between a system that tries to infer a user's nuanced, hierarchical intent and one that acts as a predictable, literal-intent engine. For a reliable database system, we argue that \textbf{predictability} is a first-class architectural property. The logical consistency of a literal-intent engine is the superior choice, as a system that does exactly what it is told is a more trustworthy foundation than one that tries to guess and frequently guesses wrong. Our work demonstrates that through deliberate prompt engineering, we can design and select for this critical property of logical consistency.
This is not to say that a specialist model like BERT~\cite{devlin19bert} could never learn to handle such nuance. Its failure in our OOD tests is likely a symptom of the data scale on which it was fine-tuned (4,000 samples). A model trained on a massive, web-scale dataset of queries and hierarchical intents might eventually learn to generalize. However, this highlights a critical practical consideration: achieving robustness in a specialist requires a significant, ongoing investment in data curation and re-training~\cite{sculley2015debt}.
In contrast, the generalist LLM offers a different paradigm: robustness through \textbf{controllability}. Our work demonstrates that through deliberate prompt engineering, we can configure the generalist to act as a predictable, literal-intent engine. This is a powerful feature: the system's logical behavior is not an emergent property of its training data, but an explicit, auditable artifact of its prompt~\cite{brown2020gpt3}. If a different logical behavior is desired (e.g., prioritizing certain constraints), one can simply \textit{tweak the prompt}, rather than curating a new dataset and re-training a model. For building reliable and maintainable database systems, we argue that this explicit controllability is a crucial architectural advantage.
\subsection{Limitations and Future Work}
Our discussion has synthesized our findings into a set of core principles and challenges for building robust natural language database interfaces. These insights, in turn, define the limitations of our current framework and chart a clear course for future work toward truly "unbound" systems.

\textbf{The Frontier of Data-Aware Linking.}
As our failure analysis on BIRD~\cite{li23bird} revealed, the challenge of \textit{data-aware} linking remains a critical open problem. While our semantics-aware SILE is a significant step forward, its inability to link based on data values within the query represents the next major frontier. This requires the system to correctly map a data value (e.g., a city name) to its corresponding column in the schema, often by disambiguating between multiple possibilities. Future work should explore efficient and scalable techniques for value-based sampling and schema indexing to bridge this gap, enabling systems to handle the full spectrum of real-world query complexities.

\textbf{Scalability of Multimodal Reasoning.}
The MMP's per-record reasoning loop, while powerful, represents a significant computational cost that is linear in the cardinality of the candidate set returned by its initial structured filter. For low-selectivity queries, this can become a prohibitive performance bottleneck. This limitation highlights the need for a more sophisticated, multi-stage query execution strategy that integrates principles from approximate query processing. Our findings in RQ1 caution against using probabilistic retrieval for the structurally-sensitive, zero-tolerance task of schema linking, where a single omission can cause catastrophic failure. However, the same technique is perfectly suited for the recall-oriented task of candidate set pruning. Future work should explore a cascaded filtering architecture where a "coarse-grained" semantic filter (e.g., a pre-computed vector index) drastically reduces the candidate set before it is passed to the "fine-grained" MMP for expensive, high-precision reasoning. This approach mirrors classic multi-stage query optimization, using cheap filters to protect expensive operators, thereby enabling scalable semantic querying over massive datasets.

\textbf{Towards Intent-Aware Interaction.}
Finally, the "Literal vs. Inferred Intent" dilemma highlighted by our RQ2 analysis underscores the need for more sophisticated, \textit{intent-aware} systems. The current framework forces a choice between a predictable but rigid literal-intent engine and a more flexible but unreliable model. The ideal system should bridge this gap. A critical direction for future research is the design of interactive architectures that can resolve ambiguity through clarification dialogues~\cite{zhao24sphinteract} (e.g., "Is the 4.5 rating a strict requirement or a preference?"). Developing the principles and mechanisms for such interactive, intent-aware query refinement is a crucial open problem for the community.

%================================================================================
\section{Related Work}
Our work is situated at the intersection of several active research areas: the evolution of Text-to-SQL systems, the specific challenge of schema linking, the emerging field of multimodal database querying, and the trend towards agentic data analysis.

\textbf{Text-to-SQL and Schema Linking.}
The challenge of schema linking has long been recognized as the "crux" of the Text-to-SQL problem~\cite{lei20reexamining}. Historically, this involved a trade-off between providing enough context and avoiding noise. However, with the advent of large-context, powerful LLMs, a new paradigm is emerging that challenges the necessity of traditional, external schema linking filters. Maamari et al.~\cite{maamari24death} argue that for state-of-the-art models, it is often safer and more effective to provide the full schema context and trust the model's internal reasoning to identify relevant elements, rather than risk an imperfect filter removing essential information.

Our work aligns with and formalizes this modern paradigm. We conceptualize this "in-context linking" as a first-class \textbf{query planning} primitive. Our Schema Introspection and Linking Engine (SILE) is a direct implementation of this principle: it leverages an LLM's reasoning over the complete schema to produce a structured query plan. Our contribution is twofold. First, we provide a concrete, reusable systems abstraction (the SILE) for this emerging approach. Second, through our rigorous comparison against a RAG baseline~\cite{lewis20rag} in RQ1, we provide strong empirical evidence for this paradigm, systematically demonstrating that older, unstructured filtering methods are architecturally prone to catastrophic \texttt{SCHEMA\_HALLUCINATION} failures. Thus, we argue not for the "death of schema linking," but for its evolution from an external filtering step into an integrated, LLM-driven planning phase.

\textbf{Multimodal Database Querying.}
The vision of querying unstructured data linked from relational tables is a growing frontier. Systems are emerging that follow two main architectural patterns. The "embed-first" approach, exemplified by Symphony~\cite{chen23symphony}, pre-builds a unified cross-modal vector index over all data assets and uses an LLM to decompose queries for execution across modalities. In contrast, the "database-first" approach of LOTUS~\cite{patel24lotus} extends the database engine with novel "semantic operators" that invoke LLMs, leveraging a query optimizer to manage costs. Our work presents a third, distinct architecture. Building on the "latent attribute" concept from GIQ~\cite{nejjar25giq}, DynaQuery's MMP uses the relational database itself as a powerful, on-the-fly filter via our "Ultimate Filtered Join" pattern. This design prioritizes adaptability to arbitrary schemas without reliance on pre-built, specialized indexes.

\textbf{LLM Agents for Data Analysis.}
A prominent recent trend is the development of LLM-powered agents for multi-step data analysis, as evaluated by benchmarks like Spider 2.0~\cite{lei24spider2}. These systems, such as RAISE~\cite{granado25raise}, decompose complex analytical tasks into an interactive, multi-turn loop of planning, tool use (e.g., SQL execution), and self-correction. This agentic paradigm is powerful but orthogonal to our work. DynaQuery focuses on perfecting the foundational, single-shot "NL-to-SQL" tool that these agents must rely on. Our contributions to the robustness and predictability of this core primitive—particularly by mitigating contextual failures like \texttt{SCHEMA\_HALLUCINATION}—can be seen as providing a more reliable foundational tool for these next-generation agentic frameworks.

\section{Conclusion}
This paper establishes a foundational architectural principle for natural language data interfaces: the treatment of schema linking is a systems-level decision that dictates robustness and generalization. We provided definitive, multi-benchmark evidence that the prevalent paradigm of unstructured RAG is architecturally brittle, prone to catastrophic \texttt{SCHEMA\_HALLUCINATION} failures. In response, we introduced \textbf{DynaQuery}, a framework built on a core tenet: schema linking must be a first-class, structure-aware query planning phase. Our \textbf{SILE} primitive embodies this principle, proving its superiority by nearly eliminating these critical errors and establishing a predictable, robust foundation.
Our work yields a conceptual roadmap for the field—the \textbf{``Hierarchy of Awareness''}—charting the path from the schema-awareness we demonstrate, to the semantics-awareness required for generalization, and ultimately to the data-awareness for truly ``unbound'' systems. By providing both a validated architectural blueprint and these core engineering principles, we bridge the critical gap between the visionary goal of universal data querying and the practical systems required to achieve it.

%================================================================================
\bibliographystyle{ACM-Reference-Format}
\bibliography{references} % This will point to our .bib file
\end{document}